\documentclass[twocolumn,groupedaddress,aps,prb,showpacs,10pt]{revtex4-1}
\usepackage[dvips]{graphicx}
\usepackage{graphicx}
\usepackage{amssymb,amsmath}
\usepackage{dcolumn}
\usepackage{bm}
\usepackage{color}

\newcommand{\oh}[1]{\textcolor{black}{ #1}}

\begin{document}

\title{The role of geometrical symmetry in thermally activated processes in clusters of interacting dipolar moments}

\author{O.~Hovorka$^{1, 3*}$, J.~Barker$^1$, G.~Friedman$^2$, R.~W.~Chantrell$^1$}
\affiliation{$^1$Department of Physics, The University of York, York, YO10 5DD, UK}
\affiliation{$^2$Electrical and Computer Engineering, Drexel University, Philadelphia, PA 19104, USA}
\affiliation{$^3$Engineering and the Environment, University of Southampton, Southampton, SO17 1BJ, UK}
\date{\today}

\begin{abstract}
\noindent
Thermally activated magnetization decay is studied in ensembles of clusters of interacting dipolar moments by applying the master-equation formalism, as a model of thermal relaxation in systems of interacting single-domain ferromagnetic particles. Solving the associated master-equation reveals a breakdown of the energy barrier picture depending on the geometrical symmetry of structures. Deviations are most pronounced for reduced symmetry and result in a strong interaction dependence of relaxation rates on the memory of system initialization. A simple two-state system description of an ensemble of clusters is developed which accounts for the observed anomalies. These results follow from a semi-analytical treatment, and are fully supported by kinetic Monte-Carlo simulations.
\end{abstract}

\maketitle

%%%%%%%%%%%%%%%%%%%%%%%%%%%%%%%%%%%%%%
\section{Introduction} Understanding the role of dipolar interactions on thermally activated processes in assemblies of magnetic nanoparticles remains a challenge despite the technological importance in many areas such as magnetic information storage~\cite{Piramanayagam}, biology and medicine~\cite{Haun2010, Pankhurst2003}. Difficulties stem from the many-body nature of the problem involving anisotropic dipolar coupling between a large number of degrees of freedom and the multivariate distribution of particle properties relevant in real systems, giving rise to a range of complex behaviors such as multi-scale dynamics~\cite{Majetich2006}, aging~\cite{Jonsson1995, Vincent2007}, spin glass phase~\cite{SpinGlYoung, Bedanta2009}, or initialization and memory effects~\cite{Sun2003, Tsoi2005, Chakraverty2005}.

Langevin type dynamics that emerge from a system of coupled stochastic Landau-Lifshitz-Gilbert equations~\cite{Coffey} have proved to be a useful approach for understanding the behavior of assemblies of interacting superparamagnetic particles and the related high frequency phenomena~\cite{Palacios1998, Berkov2001, Jonsson2001, Usadel2006, Sukhov2008}.
A numerical solution of Landau-Lifshitz-Gilbert equations requires the time step in the stochastic integration to be much smaller than the precession time, which practically limits calculations to time scales of hundreds of nanoseconds. However, thermal relaxation often takes many orders of magnitude longer.
This is particularly true for magnetically `viscous' particles in the high damping regime, in or near their blocked states. In such cases the configuration space becomes separated into virtually disconnected regions, defining discrete states of a particle system. Due to the relatively large energy barriers, the residence times in the neighborhood of these states are often very long, extending from milliseconds, for example in many biological applications, to years in magnetic recording. As a result, integrating the stochastic dynamical equations to explore the thermal relaxation dynamics in the full state space goes far beyond current computing capabilities.

Instead, a complementary framework describes the thermally activated transitions as a discrete Markov process in this state space governed by the associated master-equation (ME)~\cite{KlikJAP1994, KlikPRB1995, AmirPNAS2012}, similar to the transition state theory in chemistry~\cite{Kampen}. The problem is ideally approached by applying the kinetic (dynamic) Monte-Carlo methods which propagate the ME in time~\cite{Gillepsie1976, Chantrell2000, Fal2013}. The transition rates are taken to be of Arrhenius form, dependent on energy barriers between the available states, in analogy with the early model of N\'{e}el~\cite{Neel1949}. The crossover between the stochastic Landau-Lifshitz-Gilbert dynamics and the ME approach has been demonstrated previously in the case of ensembles of non-interacting particles~\cite{Brown1963}.

It is necessary to emphasize that such a ME framework is fundamentally different from the time-quantified Monte-Carlo approaches~\cite{NowakPRL2000, ChengPRL2006}, which inherently lack any physically motivated time scales and rely on quantifying the Monte-Carlo step by a direct association with the time normalization in the Langevin dynamics approach (or the associated Fokker-Planck equation)~\cite{MetropolisComment}. Such a `time-coarse-grained' approach does not allow resolving the spectrum of natural time scales of thermal fluctuation modes, which becomes essential if hysteresis plays a dominant role and the active part of the time scale spectrum depends on the memory of all previously visited states.

In the ME formalism employed here, such a spectrum of time scales is naturally embedded, which allows a fully resolved description of time-dependent thermal relaxation processes. This comes at a price, since the procedure generally requires the solution of a global optimization problem to obtain a topographic mapping of the entire energy landscape of an interacting particle system. This essentially means identifying all possible transition paths between the available states and the associated energy barriers $\delta e$ separating these states, determining the transition rates governing the network of probability flows within the state space. Such a procedure quickly becomes a formidable task as the system size grows.
The problem can be simplified, for instance by applying cumulant expansion methods to re-express the many-body ME problem as a hierarchy of coupled evolution equations for cumulants of perpetually increasing order, which can then be truncated to a tractable form by applying appropriate decoupling approximations~\cite{Pfeiffer1990a, Pfeiffer1990b}. Low order approximations are typically presumed, essentially neglecting any effects that might result from the correlated nature of thermal fluctuations within the interacting system.

In this way, the dipolar effect has been explored in systems of magnetic nanoparticles by varying particle concentration, clustering, and dimensionality~\cite{Morup1994, Dormann1988, Serantes2010, Toro2012, Fleutot2013, Luis2002, Poddar2002}, and revealed the enhancement or suppression of relaxation time scales in specific cases. In small clusters of nanoparticles, the finite size effects introduce a further dependence on geometry. The analyses lead to competing interpretations~\cite{Hansen1998, Dormann1999, Allia2011}, which brings into question the overall validity of the simplifying assumptions and suggests the need for a self-consistent treatment of the correlated fluctuation effects in any description of relaxation phenomena in interacting particle systems.

In this article such effects are included to a full extent to study weakly dipolar-interacting single-domain magnetic particles organized into small clusters, by adopting the full ME formalism without assuming any degree of reduction. Changing the particle cluster geometry, symmetry, and dimensionality allows a direct control of the dipolar interaction effects. Quantifying the energy landscape in terms of the barriers $\delta e$, composing the ME and solving the associated eigenvalue problem, we express the magnetization decay during the approach to equilibrium in zero external field as a weighted superposition of contributions from time scales of the available relaxation modes:
\begin{equation}\label{mdist}
    M(t) = M_0\int_0^\infty f(\epsilon) e^{-t/\tau(\epsilon)}\,d\epsilon
\end{equation}
recovering the well-known expression of Street-Woolley~\cite{Street1949, Battle1997, Kodama1999}. Here, $M_0$ is the initial magnetization, and $\tau = \tau_0\exp(\epsilon)$ following the Arrhenius law with $\tau_0$ typically taken to be a constant in the nanosecond range~\cite{Morrish, tauzero}.
As we will show, the variables $\epsilon$, being a solution of the eigenvalue problem, are distinct from the energy barriers $\delta e$, and are to be interpreted as `renormalized' or `eigen' energy barriers (in the units of thermal energy $k_BT$) representing the time scales $\tau$ of dynamical modes resulting from the correlated nature of fluctuations. These emerge even in the weak interaction limit with a likelihood of only single spin-flip events.

We show that the differences between $\epsilon$ and $\delta e$ are strongly determined by the symmetry of clusters and give rise to an initial memory dependent relaxation. Any deviations from the energy barrier picture and the memory effect disappear for symmetric spin clusters with an isotropic moment of inertia. On the other hand, in the non-symmetric cases, the same cluster structure may display both enhanced or suppressed interaction strength dependence of relaxation time scales, determined solely by the initialization prior the relaxation process, which is rather surprising. Although our approach is semi-analytical, it remains fully equivalent to kinetic Monte-Carlo modeling~\cite{Gillepsie1976, Chantrell2000, Fal2013}, as illustrated below.

\section{Theoretical approach}
We develop a semi-analytical approach, based on a master-equation formalism, applicable to weakly interacting systems.
Consider an ensemble of $q = 1,\dots, Q$ independent clusters of spins. A typical cluster $q$ contains $N_q$ interacting spins $\hat s_i$, $i=1,\dots, N_q$, represented as vectors of unit length. The governing energy density associated with the $q$-th cluster in the ensemble reads:
\begin{equation}\label{totenergy}
  e  = \sum_{i=1}^{N_q}\left((\vec k_i\times\hat s_i)^2
                                            -\hat s_i\cdot\vec h
                                            -\hat s_i\cdot\vec h_i^\textrm{dip}\right)
\end{equation}
The first term determines the preferential orientation of spins $\hat s_i$ towards their anisotropy vectors $\vec k_i$, and the remaining terms are the Zeeman energy and the dipolar interaction energy due to all neighbors of $\hat s_i$ within the cluster where $\vec h_i^\textrm{dip} = I\sum_{j\ne i}r_{ij}^{-3}(-\hat s_j +  3\hat r_{ij}(\hat s_j\cdot\hat r_{ij}))$. Here $\hat r_{ij} = \vec r_{ij}/r_{ij}$, the spin-spin separation vectors $\vec r_{ij}$ are normalized by the smallest spin-spin distance within a spin structure, $a$, and $I$ is the interaction strength. Interactions between different clusters $q$ are not considered. We use the standard phenomenology for describing thermal activation processes~\cite{SpinGlYoung, KlikPRB1995, AmirPNAS2012}, where the minima of Eq.~\eqref{totenergy}, to be denoted as $e_\alpha$, define stable configurations of the $N_q$ spins as labeled microstates $\alpha$.  At any field $\vec h$, the form of Eq.~\eqref{totenergy} forces the spins within the microstates to be bistable entities with a possibly non-collinear alignment, depending on the cluster geometry, the distribution of $\vec k_i$, and on $\vec h$. The model is analogous to a system of interacting Stoner-Wohlfarth particles~\cite{Morrish}.

The total number of microstates $\alpha = 1, \dots, \Omega_q$ of a cluster defines a discrete state-space for the stochastic thermal activation process. The rates $\tau_{\alpha\beta}$ of transitions from a microstate $\beta$ to $\alpha$ are dependent on the energy barriers $\delta e_{\alpha\beta} = e_s-e_\beta$, where $e_s$ is the saddle point energy along the transition path $\beta\rightarrow\alpha$. Throughout this work, we assume the rates to take the Arrhenius form: $\tau_{\alpha\beta}=\tau_0\exp(\delta e_{\alpha\beta})$~\cite{tauzero}.
The master-equation governing the time evolution of microstate probabilities reads~\cite{Kampen, Schnakenberg}:
\begin{equation}\label{me}
  \frac{d}{dt}p_\alpha(t) = \sum_{\beta=1}^{\Omega_q}{\cal W}_{\alpha\beta}p_{\beta}(t)
\end{equation}
including the initial condition at $t=0$, consistent with the initial microstate $\alpha_0$, and the transition matrix is ${\cal W}_{\alpha\beta}=\tau_{\alpha\beta}^{-1} - \delta_{\alpha\beta}\sum_{\gamma=1}^{\Omega_q}\tau_{\gamma\alpha}^{-1}$.

\begin{figure}[!t]
\center
\includegraphics[width=8cm, trim = 11 5 11 5]{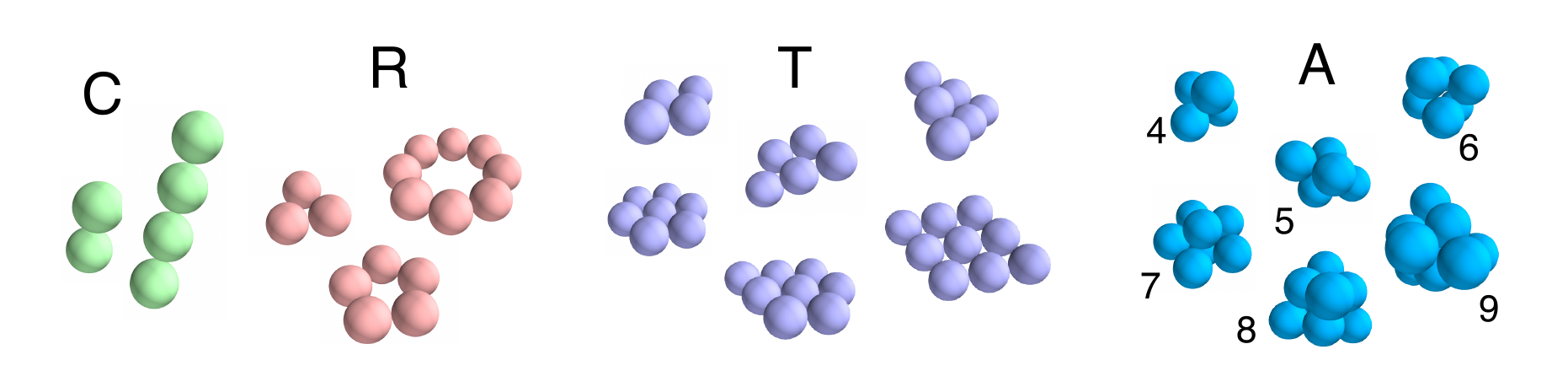}
\caption{
The cluster structures forming ensembles. (C) chains oriented along the $\hat z$-axis with $N_s = 2-9$ spins; (R) rings with $N_s=3-9$ and (T) triangles with $N_s = 4-9$ lying in the $\hat x\,\hat y$-plane; (A) 3D structures of size $N_s=4-9$ taken from Ref.~\cite{Arkus2009}.}
\label{fig:structures}
\end{figure}

If the external field $\vec h$ is constant, then both $\delta e_{\alpha\beta}$ and ${\cal W}_{\alpha\beta}$ are time invariant. The master-equation reduces to a linear system of ordinary differential equations with the most general form of solutions~\cite{Schnakenberg}:
\begin{equation}\label{me_sol}
  p_\alpha(t) = \sum_{r=1}^{\Omega_q} c_r u_\alpha^r Q_\alpha^r(t)e^{-t/\tau_r}
\end{equation}
Here $c_r$ are the initial condition-dependent constants to be found by inverting the solutions $p_\alpha$ at $t=0$, and $1/\tau_r$ and $u_\alpha^r$ are the eigenvalues and the right eigenvector components of the transition matrix, respectively.
\oh{The $Q_\alpha^r(t)$ denote the series of polynomials in time of a degree dependent on the degeneracy properties of the eigenvalue spectra~\cite{Schnakenberg}}. The numerical analysis of ensembles containing $10^6$ individual cluster structures in Fig.~\ref{fig:structures}, which will be discussed in detail below, suggests that although the eigenvalues of the transition matrices ${\cal W}$ obtained from the corresponding Eqs.~\eqref{totenergy} are often degenerate in symmetric cases, the associated right eigenvectors are always linearly independent. \oh{This therefore allows us to set $Q_\alpha^r(t)=1$ in the present study~\cite{Schnakenberg}}, which reduces Eq.~\eqref{me_sol} to a linear superposition of exponential contributions from the available relaxation time scales.
The average magnetization of a cluster is obtained from the solutions as an expectation value $M_q(t) = \sum_{\alpha=1}^{\Omega_q} m_\alpha p_\alpha(t)$, where $m_\alpha$ are the magnetizations of microstates $\alpha$ projected onto the external field vector $\hat h$: $m_\alpha = N_q^{-1}\sum_{k=1}^{N_q}\hat s_k\cdot \hat h$ and the sum runs through all $\hat s_k$ in $\alpha$. Combining the expressions and arranging gives
\begin{equation}\label{mdist_cluster}
  M_q(t) = \sum_{r=1}^{\Omega_q}\xi_r e^{-t/\tau_r(\epsilon_r)}
\end{equation}
with $\xi_r = \sum_{\alpha=1}^{\Omega_q} c_r m_\alpha u_\alpha^r$. For convenience, we also introduce the energy representation of the eigenvalues as $\epsilon_r$, related to $\tau_r$ via the Arrhenius law: $\tau_r=\tau_0\exp(\epsilon_r)$. In the following discussion, the $\epsilon_r$ will be compared to the energy barriers $\delta e_{\alpha\beta}$ obtainable by directly mapping the energy surface in Eq.~\eqref{totenergy}.

\begin{figure}[!t]
\center
\includegraphics[width=8cm, trim = 11 5 11 5]{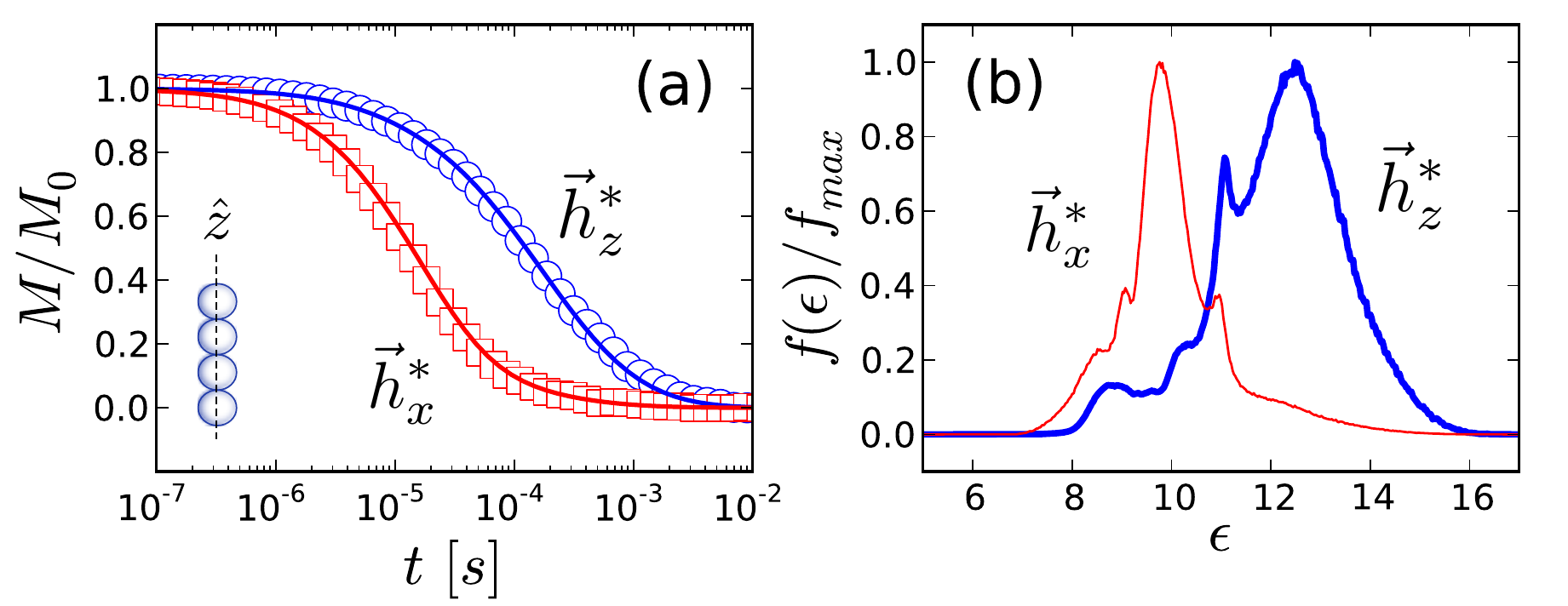}
\caption{
Relaxation in an ensemble of $10^6$ four-spin identical chains oriented along the $\hat z$-axis with a spherically random distribution of anisotropy axes: (a) Magnetization decay in external field $\vec h = \vec 0$ and (b) the corresponding $f(\epsilon)$ according to Eqs.~\eqref{mdist}-\eqref{f_epsilon} where the $\epsilon$-axis is in the units of $k_{B}T$. Initialization is in a field oriented parallel ($\vec h_z^*$) and perpendicular ($\vec h_x^*$) with respect to $\hat z$.
In (a) calculations are based on Eqs.~\eqref{mdist} and~\eqref{f_epsilon} (lines) validated by the kinetic Monte-Carlo calculations (symbols). Assumed is the model system Eq.~\eqref{totenergy} for the parameter set discussed in the text.}
\label{fig:feps}
\end{figure}

At this stage, it is useful to summarize the current notations and introduce some further notations used below. The microstate energies and the energy barriers have been distinguished by the symbols $e$ and $\delta e$, respectively. The energy equivalents of the relaxation time scales of eigenmodes $\tau$ have been denoted by $\epsilon$, and will be termed rather artistically as `eigen-barriers'. To analyze ensembles of clusters with random properties, it will also be useful to distinguish between the notations for the distribution of a property of a cluster and those of an ensemble, denoted by $\{\cdot\}_q$ and $\{\cdot\}_{q=1}^Q$ respectively.
For example, the set $\{\delta e_{\alpha\beta}\}_q$ is a distribution of energy barriers of a cluster $q$ calculated from the associated Eq.~\eqref{totenergy} and the $\{\delta e_{\alpha\beta}\}_{q=1}^Q$ a distribution of all energy barriers obtained from Eqs.~\eqref{totenergy} individually for every cluster in the ensemble. Similarly, $\{{\cal W}_{\alpha\beta}\}_q$ denotes the transition matrix associated with a cluster $q$, i.e. equivalent to ${\cal W}_{\alpha\beta}$ in the previous notation, and $\{{\cal W}_{\alpha\beta}\}_{q=1}^Q$ the ensemble of all matrices. The ensemble averages will be denoted by the triangular brackets $\langle\cdot\rangle$. To avoid confusion, where possible we will drop the subscript $\alpha\beta$ to further simplify these notations, replacing $\{{\cal W}_{\alpha\beta}\}_q$ simply by $\{{\cal W}\}_q$, etc.

The above calculation of $M_q$ in Eq.~\eqref{mdist_cluster} can be extended to find the magnetization $M$ of an ensemble, equivalent to performing the ensemble average over the distribution $\{M_q\}_{q=1}^Q$. Denoting by $\{\epsilon_r, \xi_r\}_q$, the set of pairs related via the eigenvalue problem for the $q$-th cluster, and by $\{\epsilon_r, \xi_r\}_{q=1}^Q$, the set for all clusters in the ensemble, we can generate a joint probability distribution $D(\epsilon, \xi)$ as a normalized two-dimensional histogram of $\{\epsilon_r, \xi_r\}_{q=1}^Q$. The interpretation of $\epsilon$ and $\xi$ as random variables follows in a disordered case with random parameters in Eq.~\eqref{totenergy}. The $D(\epsilon, \xi)$ reflects the fact that $\epsilon$ and $\xi$ are correlated within a spin cluster and uncorrelated between different clusters in the ensemble. The product $D(\epsilon, \xi)d\epsilon d\xi$ defines the fraction of pairs in the range $(\epsilon, \epsilon+d\epsilon)\times(\xi, \xi+d\xi)$, contributing to the overall magnetization of an ensemble by $\xi\exp(-t/\tau(\epsilon))D(\epsilon, \xi)d\epsilon d\xi$. Integrating over gives Eq.~\eqref{mdist} if:
\begin{equation}\label{f_epsilon}
  f(\epsilon) = M_0^{-1}\int_{-\infty}^{\infty}\xi D(\epsilon, \xi)d\xi
\end{equation}
where $M_0$ normalizes $f(\epsilon)$ to take the meaning of a probability density.

Thus the arguments leading to Eqs.~\eqref{mdist_cluster} and~\eqref{f_epsilon} suggest that the $\tau$ in Eq.~\eqref{mdist} are the time scales of eigenmodes $-$ correlated thermal activation modes, and thus their energy representation $\epsilon$ is generally expected to differ from the energy barriers $\delta e$ mapping the energy surface according to Eq.~\eqref{totenergy}. The coefficients $\xi$ weight the contributions of eigen-barriers $\epsilon$ to the magnetization decay. Consequently, the mean time scale of the approach to equilibrium, $\bar\tau \sim \tau_0\exp(\bar\epsilon)$, as it follows from the self-consistent pair of expressions~\eqref{mdist} and~\eqref{f_epsilon} corresponds to the mean $\bar\epsilon = \int\epsilon f(\epsilon)d\epsilon$ dependent on $\xi$ through Eq.~\eqref{f_epsilon}, rather than to the ensemble average $\langle\epsilon\rangle$ calculated over the $\{\epsilon_r\}_{i=1}^Q$.

The developed approach is fully equivalent to kinetic Monte-Carlo modeling~\cite{Chantrell2000}. This is demonstrated by the direct comparison in Fig.~\ref{fig:feps}(a), which shows magnetization decay in the external field $\vec h = \vec 0$ for two different system initializations, calculated from Eq.~\eqref{mdist} with distributions $f(\epsilon)$ obtained from Eq.~\eqref{f_epsilon} and shown in Fig.~\ref{fig:feps}(b) (lines). The agreement with the kinetic Monte-Carlo calculations (symbols) is close to exact, providing a validation of the developed master-equation formalism. The calculation details are given in the next section.

%

%%%%%%%%%%%%%%%%%%%%%%%%%%%%%%%%%%%%%%%%%%
\section{Results and Discussion}
We now apply these general considerations to investigate magnetization decay in the absence of a field ($\vec h = \vec 0$) in ensembles of the various spin cluster geometries illustrated in Fig.~\ref{fig:structures}. These include spin chains $C$ oriented along the $z$-axis of the coordinate system (1-dimensional structures), rings $R$ and triangular lattice cuts $T$ lying in the $xy$-plane (2-dimensional), and the 3-dimensional structures $A$ taken from Ref.~\cite{Arkus2009}. To emphasize the role of geometry of the spin arrangement, we will consider the case where an ensemble contains only one structure type, for example ensembles composed purely of 2-spin chains, or 5-spin rings.

Interpreting the system of Eq.~\eqref{totenergy} associated with an ensemble in terms of the Stoner-Wohlfarth model~\cite{Morrish} of spherical particles having volume $V=\pi a^3/6$ and saturation magnetization $M_s$, and normalizing the external field $\vec h$ to be in the units of energy per unit volume, the interaction strength reads $I = \mu_0M_s^2/3$. The shortest spin-spin distance for all structures is chosen to be $a$, i.e. the nearest neighbor particles being in contact.
We assume a practically relevant case where the anisotropy vectors $\vec k_i$ in Eq.~\eqref{mdist} are randomly distributed (uniform distribution on a sphere), and for simplicity take $|\vec k_i| = k$ for all $i$. Such a choice of the distribution results in no preferential anisotropy orientation in an ensemble and therefore if relaxation occurs in $\vec h = 0$ as is the case here, the only symmetry breaking element in the description of an ensemble by the system of Eqs.~\eqref{totenergy} may be the spin cluster geometry. It is then expected that the symmetry is always broken for the ensembles of spin-chains, as opposed to ensembles of clusters with higher geometrical symmetries such as that of the pyramidal structure.

While the conclusions below are general and based on thorough testing, the specific parameter set used in simulations here was $a = 10$ nm, $T = 300$ K, $|\vec k| = 10^7$ J/m$^3$, $M_s\le 4\times 10^5$ A/m ($\sim$Fe$_3$O$_4$ particles), giving $KV/k_BT\approx 12.5$ and the maximum interactions strength $I(M_s = 4\times 10^5 \,\textrm{A/m}) = I_0 = 67.02\times 10^3$ J/m$^3$, which will be used as a reference. Thus $|\vec k| >> I_0$, consistent with the assumption of weak interactions. In the present study, all ensembles are generated to consist of up to $10^6$ spin cluster structures. The calculations are carried out as follows.

\emph{1. Assembling the transition matrix $\{{\cal W}\}_q$ for the $q$-th cluster.}
Setting $\vec h = \vec 0$ during the magnetization thermal decay process implies a time-invariant energy landscape, and the set of microstate energies $\{e_\alpha\}_q$ associated with the $q$-th cluster can be identified as local minima of Eq.~\eqref{totenergy} written for the $q$-th cluster. Determining all available local minima is generally a difficult task requiring sophisticated minimization procedures which soon becomes intractable as the cluster size $N_s$ grows~\cite{Franco2011, Bessarab2013}. The problem simplifies in the weak interaction limit, which implies: 1) the energy hypersurface is a smooth deformation of the noninteracting case and 2) a likelihood of single-spin transitions only.
Then all microstate energies $e_\alpha$ can be identified by consecutively choosing the microstates of the non-interacting case as initialization, and for every such choice individually applying the iterative scheme based on: (i) rotating every spin $\hat s_i$ within the selected microstate to a new orientation consistent with interactions: $\hat s_i' = \hat s_i + \gamma(-\partial e/\partial\hat s_i-\hat s_i)$, where the derivative is the effective field acting on $\hat s_i$ and $\gamma$ is a convergence criterion, and (ii) checking if the error $\sum_i |\hat s_i'-\hat s_i| < \textrm{tolerance}$. If the tolerance condition has been achieved, selecting the next microstate, otherwise repeating (i)-(ii) (we set $\gamma = 0.55$ and tolerance = $10^{-4}$).
Given the weak interaction limit, this procedure results only in a smooth adjustment of the non-interacting spin components used for initialization into valid microstates with energies $e_\alpha$ and magnetizations $m_\alpha$ consistent with the interaction structure of the cluster.
Subsequently, the energy barriers $\{\delta e_{\alpha\beta}\}_q$ for single-spin transitions in the cluster's state space can be identified by selecting the pairs of microstates $\alpha$ and $\beta$ related by only one reversed spin $j$, which relates to the $j$-th term in Eq.~\eqref{totenergy}. The barrier $\delta e_{\alpha\beta}$ along the transition path $\beta\rightarrow\alpha$ associated with the switching of the $j$-th spin then equals the difference $e_s-e_\beta$, with $e_s$ being the maximum of the $j$-th term in Eq.~\eqref{totenergy}~\cite{ebapprox}. Finally, having obtained the set of barriers $\{\delta e_{\alpha\beta}\}_q$ allows to define the transition matrix $\{{\cal W}_{\alpha\beta}\}_q$ by the element-wise application of the Arrhenius law.
Similarly, the microstate magnetizations $\{m_\alpha\}_q$ required for the evaluation of the weighting coefficients $\xi$ can be obtained from the spin patterns within the microstates $\alpha$ by following the recipe leading to Eq.~\eqref{mdist_cluster}.

\emph{2. Initialization and calculation of $f(\epsilon)$.}
The transition matrix $\{{\cal W}_{\alpha\beta}\}_q$ fully characterizes the thermal relaxation process of the $q$-th cluster for $t>0$ only after specifying an initial condition. To generate the initial condition we imitate the standard initialization procedure applied during $t<0$ by rapidly reducing the saturating external field $\vec h^*$ to a final value $\vec h^*=\vec h=\vec 0$ attained at $t=0$, and holding it fixed afterwards during the decay process. Note the different notations used for the initializing field, $\vec h^*$, and the field $\vec h$ applied during the magnetization decay. This field history is here modeled as an athermal rate-independent hysteresis process~\cite{Berkov1996}, essentially by minimizing the system energy at every field step, and initializes the cluster in the microstate $\alpha_0$ which implies $p_\alpha(0)=1$ for $\alpha = \alpha_0$ and $p_\alpha(0)=0$ otherwise. Thus the microstate $\alpha_0$ is consistent with the interaction structure of the cluster and with the field history $\vec h^*$ applied at a sufficiently fast rate for thermal fluctuations to be irrelevant.
Having obtained the initial condition and using the standard numerical techniques~\cite{nrc} to solve the eigenvalue problem for $\{{\cal W_{\alpha\beta}}\}_q$ obtained in \emph{1.} above, allows the determination of coefficients $\{c_r\}_q$ by inverting the solutions for $p_\alpha(0)$ in Eq.~\eqref{me_sol}, identifying the sets $\{\epsilon_r\}_q$ and $\{\xi_r\}_q$ according to the discussion of Eq.~\eqref{mdist_cluster} and generating the combined set $\{\epsilon_r, \xi_r\}_q$.

Repeating the above procedures \emph{1.} and \emph{2.} for every cluster in the ensemble in turn generates the ensemble of matrices $\{{\cal W}_{\alpha\beta}\}_{i=0}^Q$ and initial microstates $\{\alpha_0\}_{i=0}^Q$ giving the full set of pairs $\{\epsilon_r, \xi_r\}_{i=1}^Q$, which after histogramming, produces the joint probability distribution $D(\epsilon, \xi)$. Then $f(\epsilon)$ is computed by evaluating the integral in Eq.~\eqref{f_epsilon} and, subsequently, the magnetization decay follows from Eq.~\eqref{mdist}.
\begin{figure}[!t]
\center
\includegraphics[width=8cm, trim = 11 5 11 5]{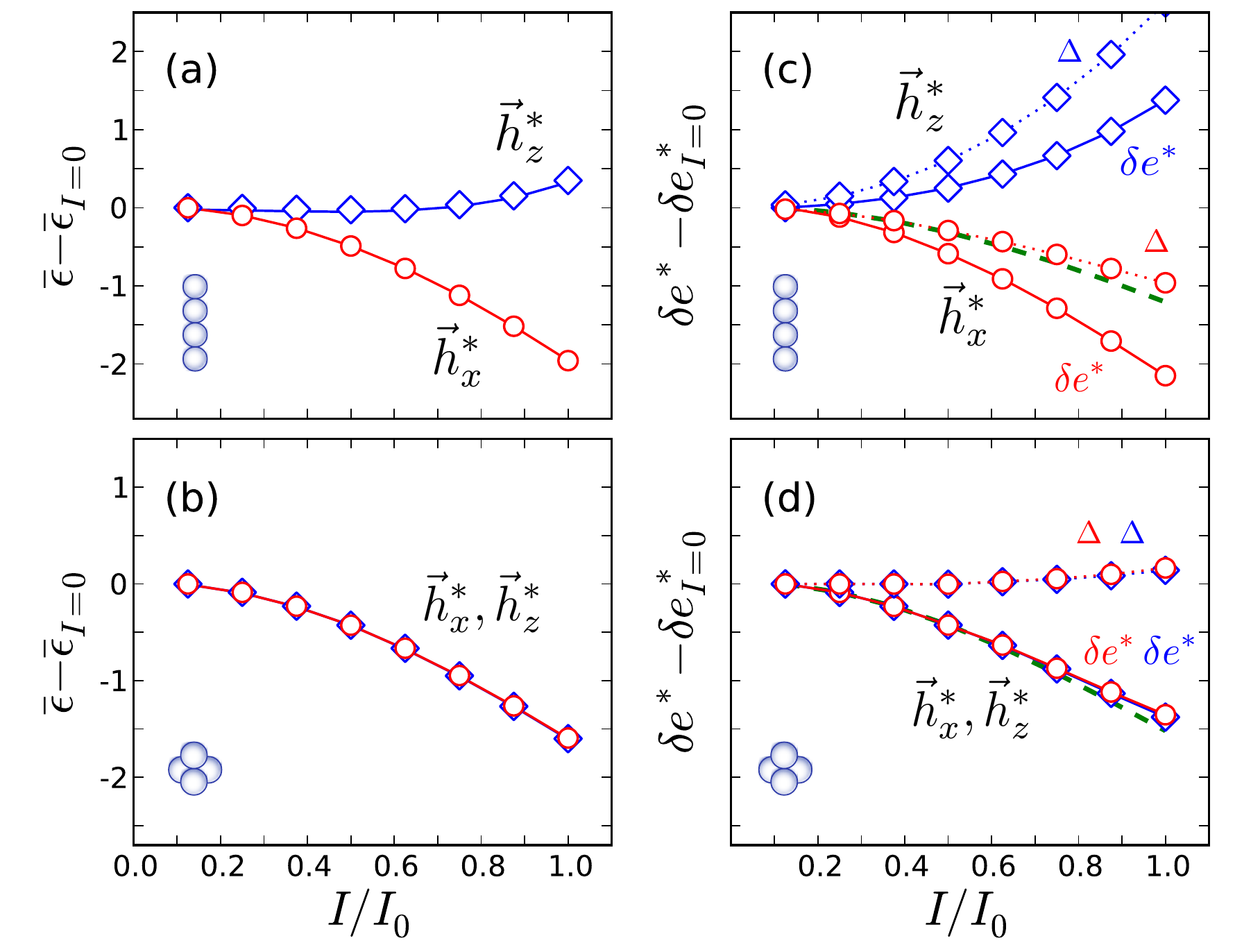}
\caption{
Dipolar interaction dependence of: the mean eigen-barrier $\bar\epsilon=\int\epsilon f(\epsilon)d\,\epsilon$ (relative to the noninteracting case $I=0$) for an ensemble of (a) 4-spin chains as in Fig.\ref{fig:feps} and (b) clusters of 4-spins arranged into an equilateral triangle-based pyramid; and of the initializaton dependent energy barrier $\delta e^*$ and correction $\Delta$ defined in Eq.~\eqref{eb_empirical} for ensembles of (c) 4-spin chains and (d) pyramids in (a)-(b). Initialization is in $\vec h_z^*$ ($\diamond$) and $\vec h_x^*$ ($\circ$), $I_0$ defined in the text. The solid lines in (a)-(b) correspond to the fits by Eq.~\eqref{eb_empirical}. In (c) and (d), the solid and dotted lines are only guiding lines while the dashed lines corresponds to the mean energy barrier $\langle\delta e\rangle$ obtained over the entire ensemble.}
\label{fig:chains_int}
\end{figure}

The application of the approach to an ensemble of 4-spin chains is demonstrated in Figs.~\ref{fig:feps}(a)-(b) and~\ref{fig:chains_int}(a). Fig.~\ref{fig:feps}(a) confirms the full consistency of the developed approach with the kinetic Monte-Carlo simulations~\cite{Chantrell2000}. The rate of magnetization decay depends on initialization, such as in the field parallel $(\vec h_z^*)$ and perpendicular $(\vec h_x^*)$ with respect to the $z$-axis, which is also reflected by the shift of $f(\epsilon)$ in Fig.~\ref{fig:feps}(b).
The initialization also influences the interaction dependence of $f(\epsilon)$, as shown in Fig.~\ref{fig:chains_int}(a) by either the increasing or decreasing trend of the mean eigen-barrier $\bar\epsilon$. Interestingly, similar differences seem practically absent for a symmetric spin structure in Fig.~\ref{fig:chains_int}(b).
These observations suggest that without specifying the initialization protocol, the interaction dependence of the eigen-barriers $\epsilon$ and of the system energy barriers $\delta e$ may be non-unique, depending on the structure type.

The next goal is to quantify the observed behavior by developing a simple phenomenological picture relating the mean $\bar\epsilon$, which effectively determines the mean relaxation time scale of the approach to equilibrium $\bar\tau=\tau_0\exp(\bar\epsilon)$, to the overall distribution of energy barriers $\{\delta e_{\alpha\beta}\}_{i=1}^Q$ in the ensemble, such that it includes the dependence on the initialization.

\begin{figure}[!t]
\center
\includegraphics[width=8cm, trim = 11 5 11 5]{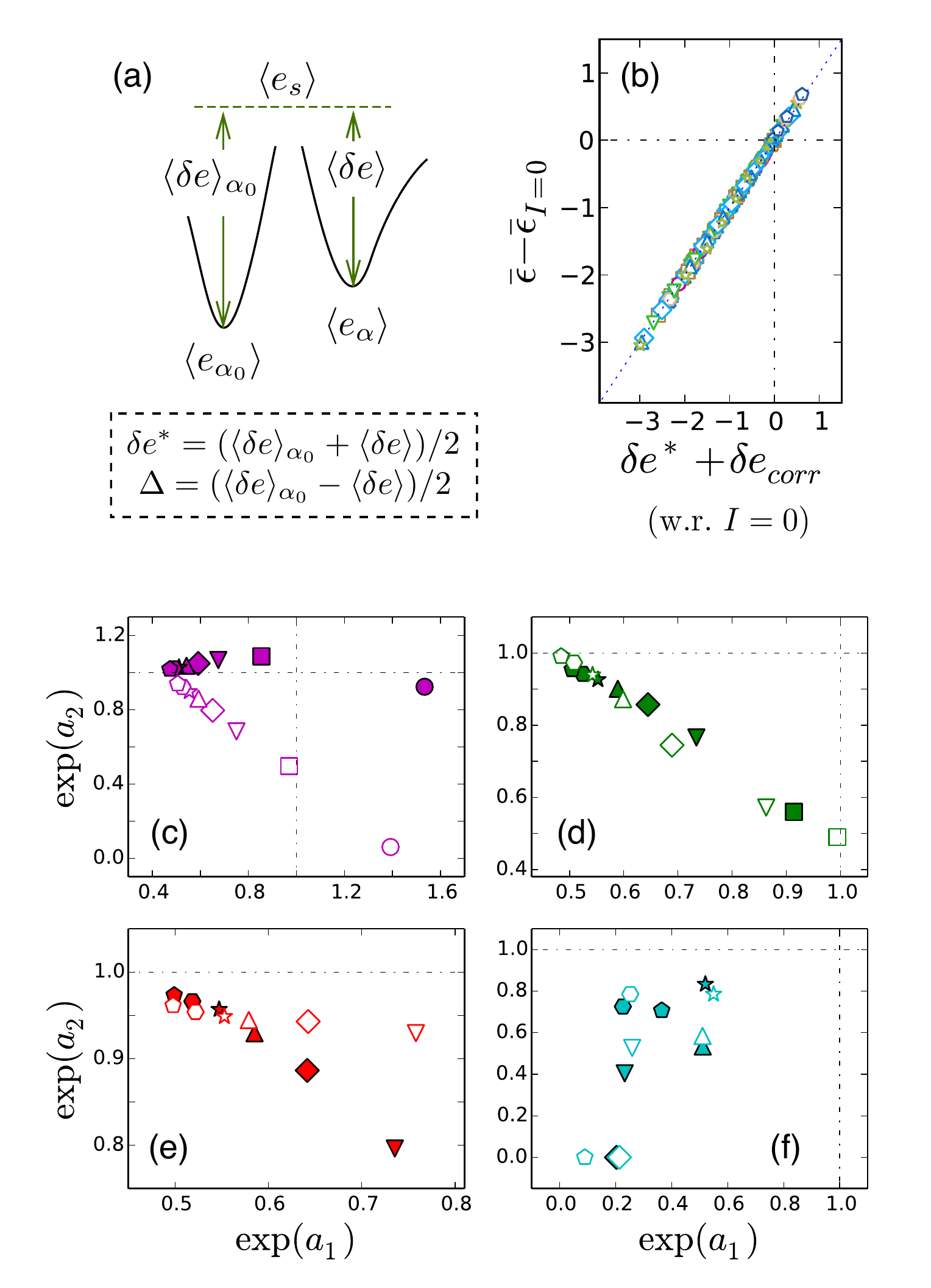}
\caption{Definition and validation of Eq.~\eqref{eb_empirical}. (a) The two-state system model of an ensemble which incorporates the dependence on initialization and leads to Eq.~\eqref{eb_empirical}. (b) The data collapse generated by fitting Eq.~\eqref{eb_empirical} to all ensembles of structures in Figs.~\ref{fig:structures} (total 512 points), validating Eq.~\eqref{eb_empirical}. Subfigures (c), (d), (e) and (f) show exponential plots of the fit coefficients $a_1$ and $a_2$ for ensembles of structures $C$, $R$, $T$ and $A$, respectively. The filled and open symbols relate to initialization in $\vec h_z^*$ and $\vec h_x^*$, and $N_s=2\,(\circ),\,3\,(\square),\,4\,(\triangledown),\,5\,(\diamond),\,6\,(\triangle),\,7\,(\star),\,8\,(\textrm{hexa}),\,9\,(\textrm{penta})$.}
\label{fig:two_level}
\end{figure}

The effect of initialization can be intuitively understood as being a result of a system following different paths along the energy landscape during its time evolution from different initial states~\cite{Oscar2004}.
The initialization procedure \emph{2.} outlined above produces a distribution of initial microstates of clusters, $\{\alpha_0\}_{q=1}^Q$, which determines the origin of probability flow in the state-space of an ensemble.
The mean time scale of the approach to equilibrium $\bar\tau\sim\tau_0\exp(\bar\epsilon)$ is expected to depend on the ensemble average $\langle\delta e\rangle_{\alpha_0}$ of restricted energy barriers $\{\delta e_{\alpha\alpha_0}\}_{i=1}^Q$ surrounding the initial microstates $\{\alpha_0\}_{q=1}^Q$. In this sense $\langle\delta e\rangle_{\alpha_0}$ relates to the `forward' probability flow away from the initial state.
In addition, $\bar\tau$ depends also on the `backward' probability flow determined by the ensemble average $\langle\delta e\rangle$ obtained over the full energy barrier distribution $\{\delta e_{\alpha\beta}\}_{i=1}^Q$, which relative to $\langle\delta e\rangle_{\alpha_0}$ gives a measure of inhomogeneity of the overall energy landscape.
This suggests that in the first order approach the $\bar\tau$ may be seen as a result of superposition of these competing probability flows, in analogy with a fictitious two-level system illustrated in Fig. ~\ref{fig:two_level}(a).

Thus, to include the initial condition dependence, we simply express within a coarse-grained description the average energy barrier surrounding the initial state as $\langle\delta e\rangle_{\alpha_0} = \langle e_s\rangle - \langle e_{\alpha_0}\rangle$ and the mean energy barrier of the ensemble as $\langle\delta e\rangle = \langle e_s\rangle - \langle e_\alpha\rangle$ as illustrated in Fig.~\ref{fig:two_level}(a), where the ensemble averages $\langle e_{\alpha_0}\rangle$, $\langle e_\alpha\rangle$, and $\langle e_s\rangle$ are to be taken over the distributions of energies of initial microstates $\{e_{\alpha_0}\}_{i=1}^Q$, all microstates $\{e_\alpha\}_{i=1}^Q$ and saddles along the transition paths $\{e_s\}_{i=1}^Q$, respectively.
It is convenient to redefine the two-state system variables by introducing the mean energy barrier $\delta e^* = (\langle\delta e\rangle_{\alpha_0}+\langle\delta e\rangle)/2$ and the mean difference $\Delta = (\langle\delta e\rangle_{\alpha_0}-\langle\delta e\rangle)/2$, noting that $\delta e^*$ incorporates the dependence on the initial state as determined by the field history $\vec h^*$  and that $\Delta$ effectively relates to the inhomogeneity of the energy hypersurface. Next we express the mean eigen-barrier $\bar\epsilon$ in terms of the energy barrier equivalents $\delta e^*$ and $\Delta$ as:
\begin{equation}\label{eb_empirical}
\bar\epsilon=\delta e^* + a_1\Delta + a_2\Delta^2
                                       =\delta e^* + \delta e_\textrm{corr}
\end{equation}
where the correction term reads $\delta e_\textrm{corr} = a_1\Delta + a_2\Delta^2$. The empirical coefficients $a_1$ and $a_2$ are to be identified by fitting Eq.~\eqref{eb_empirical} to the interaction strength dependence of $\bar\epsilon$ vs. $(\delta e^*, \Delta)$ for a given ensemble.
Examples of such fits are shown by lines in Figs.~\ref{fig:chains_int}(a) and (b) involving data in Figs.~\ref{fig:chains_int}(c) and (d), respectively. Thus $a_1$ and $a_2$ are no longer expected to depend on the interaction strength $I$ explicitly, they are however generally dependent on interactions through various structural factors such as the geometry of arrangement of spins within clusters, anisotropy and volume distributions, etc. Since the present study associates the spins with a uniform volume and assumes a spherical distribution of anisotropy axis, the $a_1$ and $a_2$ and thus the $\delta e_\textrm{corr}$ are dependent only on the spin cluster geometry.

Quantitative validation of Eq.~\eqref{eb_empirical} is shown in Fig.~\ref{fig:two_level}(b), where 512 ensembles of various spin structures $C$, $T$, $R$, $A$ listed in Fig.~\ref{fig:structures} are simultaneously fitted for different $N_s$ and initializations, giving a perfect linear data collapse. This validates the equality sign in Eq.~\eqref{eb_empirical}. For completeness, the coefficients $a_1$ and $a_2$ obtained from the fits are summarized in the exponential plots in Figs.~\ref{fig:two_level}(c)-(f) and are clearly dependent on the cluster structure and on initialization.

The correction $\delta e_\textrm{corr}$ in Eq.~\eqref{eb_empirical} is a measure of the difference between the mean eigen-barrier $\bar\epsilon$ and the mean energy barrier $\delta e^*$ obtainable directly from the topography of the energy surface. In this sense, the $\delta e_\textrm{corr}$ quantifies the validity of the energy barrier picture in the quantitative description of the approach to equilibrium by Eq.~\eqref{mdist}, as opposed to the need for the full solution $\bar\epsilon$ evaluated by solving the master-equation.
Fig.~\ref{fig:chains_int} suggests that, although the $\delta e_\textrm{corr}$ may not always be negligible, the interaction dependence of $\delta e^*$ in Fig.~\ref{fig:chains_int}(c)-(d) qualitatively resembles the trends of $\bar\epsilon$ in Figs.~\ref{fig:chains_int}(a)-(b) in both the case of ensembles of chains and of pyramids, initialized in the perpendicular field cases $\vec h_z^*$ and $\vec h_x^*$.
In Fig.~\ref{fig:chains_int}(c) the behavior of $\delta e^*$ qualitatively captures the increasing and decreasing interaction trends of $\bar\epsilon$ corresponding to the different initializations. On the other hand, in Fig.~\ref{fig:chains_int}(d) it turns out that because $\Delta\approx 0$, the $\bar\epsilon\approx\delta e^*$, implying $\langle\delta e\rangle_{\alpha_0}\approx\langle\delta e\rangle$ and indicating a relative homogeneity of the energy landscape.
For illustration, we also added the mean energy barrier $\langle\delta e\rangle$ as dashed line in Figs.~\ref{fig:chains_int}(c)-(d), which is independent of initialization since the overall energy landscape does not change with time during relaxation.

\begin{figure}[!t]
\center
\includegraphics[width=8cm, trim = 11 5 11 5]{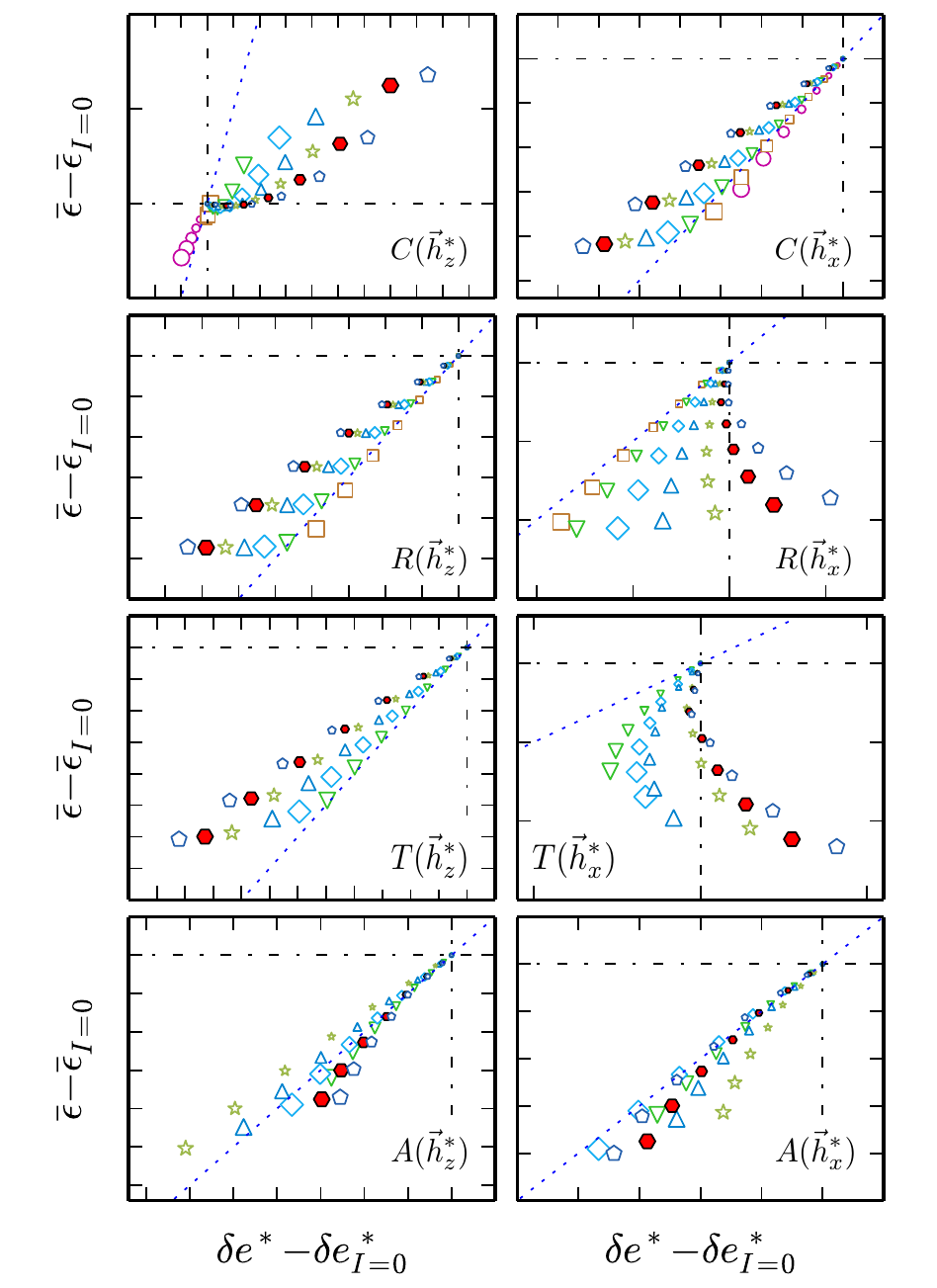}
\caption{The relative $\bar\epsilon$ vs. $\delta e^*$ for ensembles of $10^6$ individual structures in Fig.~\ref{fig:structures} and initializations in $\vec h_z^*$ and $\vec h_x^*$. $N_s=2\,(\circ),\,3\,(\square),\,4\,(\triangledown),\,5\,(\diamond),\,6\,(\triangle),\,7\,(\star),\,8\,(\textrm{hexa}$, filled symbols$),\,9\,(\textrm{penta})$ and the symbol size grows with the increasing interaction strength $I$. In every subplot the axis division unit equals 0.5. The guide lines: $\bar\epsilon$=$\delta e^*$ (dotted) and $\bar\epsilon = \delta e^* = 0$ (dash-dotted).}
%\label{fig:struct_de}
\label{fig:struct_de}
\end{figure}

The qualitative differences between $\bar\epsilon$ and $\delta e^*$ are studied systematically in Fig.~\ref{fig:struct_de} which compares $\bar\epsilon$ vs. $\delta e^*$ for ensembles of clusters of various geometries listed in Fig.~\ref{fig:structures}, initializations in $\vec h_z^*$ and $\vec h_x^*$, and all values of interaction strengths $I$ as in Fig.~\ref{fig:chains_int}. The dash-dotted lines in every subfigure are the coordinate system and the dotted line is the equality line $\bar\epsilon = \delta e^*$. Thus deviations of symbols from this line relate directly to $\delta e_\textrm{corr}$ and quantify the validity of the energy barrier picture. The axis division unit is 0.5 in all cases.

Subplot $C(\vec h_z^*)$ shows the behavior for ensembles of chains initialized in $\vec h_z^*$. The different kinds of symbols correspond to spin chains of different lengths $N_s$ as listed in the figure caption and the growing symbol size signifies the increasing interaction strength $I$. The majority of the data points are located in the upper half of the coordinate system which, given the direction of the symbol size increase, indicates increasing trends of both $\bar\epsilon$ and $\delta e^*$ vs. $I$. For the ensemble of 2-spin chains (circles) the trend is decreasing.
On the other hand, subplot $C(\vec h_x^*)$ shows behavior for ensembles of chains initialized in the perpendicular field $\vec h_x^*$, where both $\bar\epsilon$ and $\delta e^*$ decrease with the increasing $I$ for all $N_s$. Thus given that these cases of thermal decay in the ensembles $C(\vec h_z^*)$ and $C(\vec h_x^*)$ differ only by the initial condition, due to the choices of spherical anisotropy distribution and of setting $\vec h = 0$ during relaxation, this clearly shows that initialization may significantly influence the behavior and result in qualitatively different dependencies as a function of the interaction strength. It may also be noticed that in both subfigures the deviations $\delta e_\textrm{corr}$ become more pronounced as the spin chain size $N_s$ increases.

The situation is similar for ensembles of spins arranged into rings $R$, triangles $T$, and 3-dimensional structures $A$. The interaction dependent trends of $\bar\epsilon$ and $\delta e^*$ are decreasing and in mutual qualitative agreement, as again manifested by the data points lying either in the first or in the third quadrant of the coordinate system and the interaction strength increase in the direction away from the coordinate system origin. A few exceptions emerge for ensembles of structures $T(\vec h_x^*)$ and $R(\vec h_x^*)$ as $N_s$ grows; the behavior for rings $R(\vec h_x^*)$ resembles that of triangular structures $T(\vec h_x^*)$ if $N_s$ is small, however, as $N_s$ increases the geometry of rings gradually begins to effectively approach that of chains, which leads to the observed crossover $R(\vec h_x^*)\rightarrow C(\vec h_z^*)$ through the second quadrant.
Furthermore, it is a common feature in the ensembles $C$, $R$, and $T$ that increasing the cluster size gives rise to the systematic increase of the correction $\delta e_\textrm{corr}$, i.e. more pronounced deviations from the energy barrier picture, and that this occurs for both types of initializations. However, in some cases of structures of type $A$ the $\delta e_\textrm{corr}$ no longer increases monotonically with $N_s$. For example, the ensemble of 4-spin clusters $A$ displays larger $\delta e_\textrm{corr}$ than the ensemble of 5-spin clusters. In addition, in some cases of structures $A$ with higher geometrical symmetry, the observed $\delta e_\textrm{corr}$ is small and the effect of initialization negligible.

\begin{figure}[!t]
\center
\includegraphics[width=8cm, trim = 11 5 11 5]{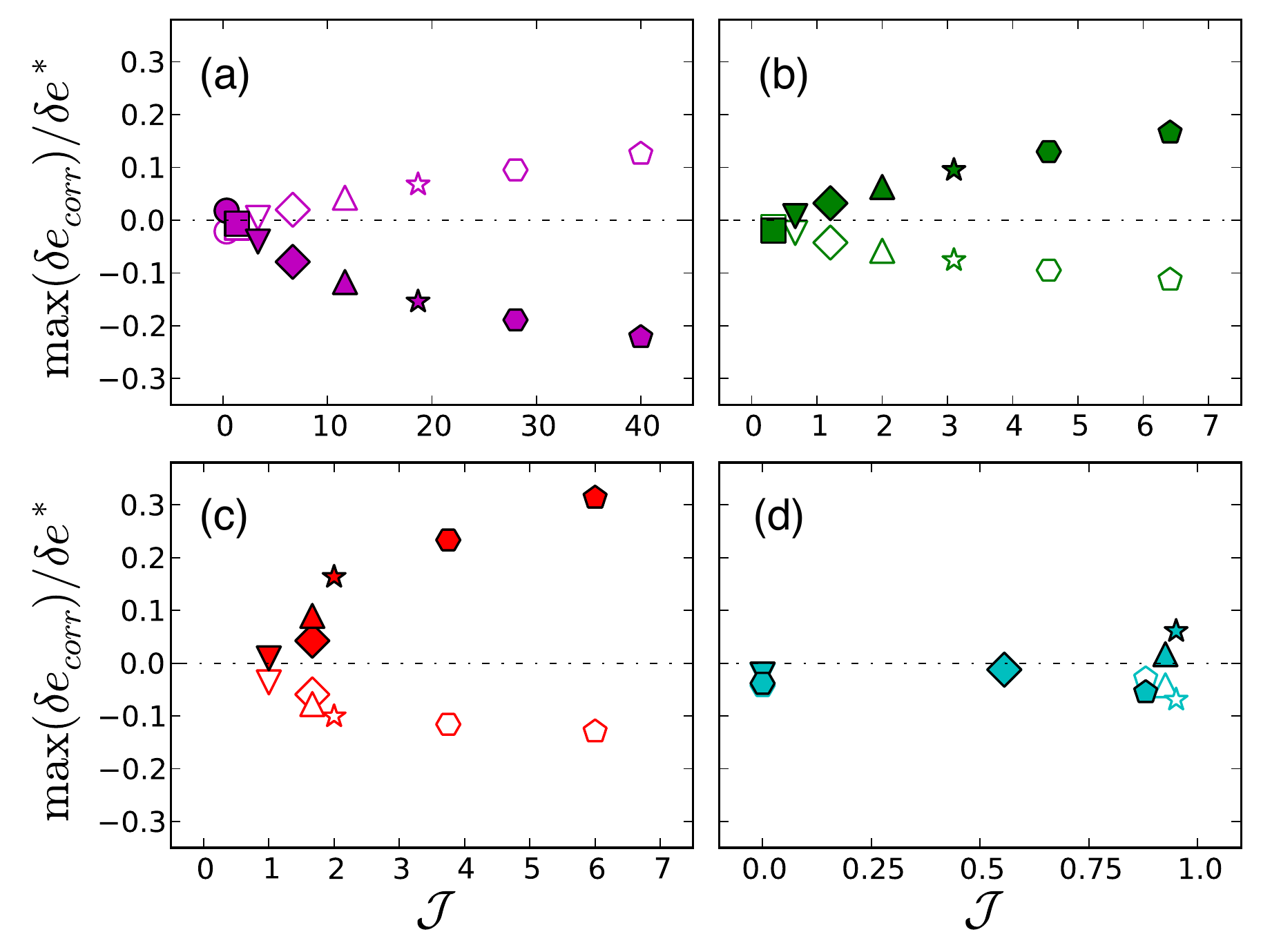}
\caption{
The upper estimates of the relative $\max(\delta e_\textrm{corr})/\delta e^*$ as a function of the asymmetry measure ${\cal J}$ for the cluster types $C$, $R$, $T$, and $A$ shown in (a), (b), (c), and (d), respectively. The filled and open symbols relate to initialization in $\vec h_z^*$ and $\vec h_x^*$, and $N_s=2\,(\circ),\,3\,(\square),\,4\,(\triangledown),\,5\,(\diamond),\,6\,(\triangle),\,7\,(\star),\,8\,(\textrm{hexa}),\,9\,(\textrm{penta})$.}
\label{fig:struct_fits}
\end{figure}

These observations suggest the role of cluster geometry in the quantitative description by Eq.~\eqref{eb_empirical} and possibly a relation to the dimensionality of a system. To check this, we define as a measure of symmetry a cumulative sum of the differences of the principal moments of inertia obtained by combining the eigenvalues ${\cal J}_{xx}$, ${\cal J}_{yy}$, ${\cal J}_{zz}$ of the cluster's moment of inertia tensor: ${\cal J} = |{\cal J}_{xx}-{\cal J}_{yy}|+|{\cal J}_{xx}-{\cal J}_{zz}|+|{\cal J}_{yy}-{\cal J}_{zz}|$. Thus ${\cal J}\rightarrow 0$ for a spherically symmetric structure, and ${\cal J} > 0$ for a structure with anisotropic geometry.
In Fig.~\ref{fig:struct_fits}, the $\max(\delta e_\textrm{corr})/\delta e^*$ vs. ${\cal J}$ is shown for the ensembles of spin clusters $C$, $T$, $R$, and $A$ of varying $N_s$ and initializations. The $\max(\delta e_\textrm{corr})$ corresponds to the upper estimate of the correction, i.e. the maximum $|\delta e_\textrm{corr}|$ from all $I$ for a given structure type, and equals the maximum deviation from the dotted lines in Fig.~\ref{fig:struct_de}. The systematic increase of the relative correction with increasing ${\cal J}$, and thus the breakdown of the energy barrier picture, is clearly demonstrated for all types of structures considered and correlates well with the dependence on initialization. For example, the highly symmetric structures $N_s = 4, 5, 8$ in $(A)$ as well as the small size structures $C$, $R$, and $T$ show practically no memory of initialization during the magnetization decay in the approach to equilibrium. Thus, this confirms the fundamental relation between the geometry of spin arrangements, initialization dependence of relaxation, and the validity of the energy barrier picture.

\section{Conclusion}
As a main conclusion, the developed ME framework allows the quantification of the validity of the conventional energy barrier picture which is widely used for interpreting experimental and computational studies of the relaxation behavior in magnetic nanoparticle systems. It shows, that the energy barrier picture neglects important aspects of the correlated nature of thermal fluctuations, and as a result cannot reproduce the initialization dependence, effects of geometry, symmetry properties, or dimensionality of interacting structures on thermal relaxation processes. This implies that the inverse problems to quantify dipolar interactions from experiments are ill-posed, where, for example, the same structures may display both increasing or decreasing interaction trends of relaxation time scales (i.e. the approach to equilibrium), depending solely on the character of sample preparation prior the relaxation process.

As has been shown, the main reason for the breakdown of the energy barrier picture is the dynamical character of thermal activation as a random walk in a spatially distributed energy landscape which, due to the correlations resulting from spatial inhomogeneities in the energy space, renormalizes the energy barriers $\delta e$ to eigen-barriers $\epsilon$ consistent with the probabilistic ME dynamics. Only a relative spatial homogeneity of the energy landscape emerging in symmetric structures preserves the validity of the energy barrier picture.
To reconcile the discrepancies, we developed a simple two-state system description introducing the notion of the initial condition dependent energy barrier $\delta e^*$, which qualitatively captures the behavior of the eigen-barriers $\epsilon$. Future work will also explore the effect of non-zero applied magnetic field on the magnetization decay, which is expected to act as a symmetry breaking element controlling the uniformity of the energy landscape and thus, according to the present study, also the applicability of the energy barrier picture.

Although the present study could not be extended to bulk systems due to the computational costs, the converging trends with the increasing size $N_s$ seen in Fig.~\ref{fig:struct_fits} suggest that similar behavior may persist even towards the bulk size, at least in structures with reduced dimensionality.
Our study is directly relevant to experimental magnetorelaxometry, which is a basis for biological sensing methodologies~\cite{BriantJMMM2011, BriantJMMM2012} and in the emerging research field of magnetic particle imaging (MPI)~\cite{Wiekhorst2012}. Furthermore, our findings are also fundamental to interpreting the rate-dependent experiments, such as the field or temperature dependent magnetization or susceptibility measurements, where the blocking temperature dependencies are typically quantified by assuming the energy barrier picture~\cite{Morup1994, Dormann1988, Serantes2010, Toro2012, Fleutot2013, Luis2002, Poddar2002, Hansen1998, Dormann1999, Allia2011}. In such cases the description is however more involved because the correlated thermal fluctuation effects further compete with the time scales of external driving forces, which need to be included in the mathematical framework if the full physical interpretation is to be acquired.

%%%%%%%%%%%%%%%%%%%%%%%%%%%%%%%%%%%%%%
The authors would like to thank A. Amir, M. Gmitra, T. Fal, O. Chubykalo-Fesenko, \`O. Iglesias, and M. Dimian for stimulating discussions. OH gratefully acknowledges support from a Marie Curie Intra European Fellowship within the 7th European Community Framework Programme under grant agreement PIEF-GA-2010-273014.

%%%%%%%%%%%%%%%%%%%%%%%%%%%%%%%%%%%%%%


\begin{thebibliography}{11}
\bibitem{MetropolisComment} The time-quantified Monte-Carlo methods typically use the Metropolis algorithm based on accepting/rejecting a trial move, which generates a random walk in the energy space that can also be represented by a master-equation. However, this equation lacks a physical time scale and is fundamentally different from the master-equation approach considered throughout this work.
\bibitem{tauzero}We note that $\tau_0$ generally depends both on external and on interaction fields. However, because this work considers relaxation in zero external field and the weak dipolar interaction limit, where the interaction strength is negligible with respect to the mean energy barrier height, approximating $\tau_0$ by a constant seems reasonable.
\bibitem{ebapprox} Note that this is an approximation expected to be valid only in the weakly interacting case.
\end{thebibliography}
\end{document}